\documentclass[aps,pre,twocolumn,superscriptaddress]{revtex4-1}

\usepackage{xcolor,hyperref}
\hypersetup{
   colorlinks,
   linkcolor={blue!50!black},
   citecolor={blue!50!black},
   urlcolor={blue!80!black}
}

\usepackage{amsmath}
\usepackage{amssymb}
\usepackage{color}

\usepackage{graphicx}

\DeclareMathAlphabet{\mathitbf}{OML}{cmm}{b}{it}

\newcommand{\zerovector}{\mathBold 0}

\newcommand{\nv}{\mathitbf n}

\newcommand{\xv}{\mathitbf x}
\newcommand{\dv}{\mathitbf d}
\newcommand{\Xiv}{{\bf\Xi}}
\newcommand{\calBold}[1]{\mbox{\boldmath${\cal #1}$}}
\newcommand{\mathBold}[1]{\mbox{\boldmath$#1$}}
\newcommand{\dbar}{{\,\mathchar'26\mkern-12mu d}}

\newcommand{\sFrac}[2]{{\textstyle\frac{#1}{#2}}}

\begin{document}

\title{Characterizing nonaffinity upon decompression of soft-sphere packings}

\author{Stefan Kooij}
\affiliation{Van der Waals-Zeeman Institute, University of Amsterdam, Science Park 904, Amsterdam, Netherlands}
\author{Edan Lerner}
\affiliation{Institute for Theoretical Physics, University of Amsterdam, Science Park 904, Amsterdam, Netherlands}

\begin{abstract}
\noindent
Athermal elastic moduli of soft sphere packings are known to exhibit universal scaling properties near the unjamming point, most notably the vanishing of the shear-to-bulk moduli ratio $G/B$ upon decompression. Interestingly, the smallness of $G/B$ stems from the large nonaffinity of deformation-induced displacements under shear strains, compared to insignificant nonaffinity of displacements under compressive strains. In this work we show using numerical simulations that the relative weights of the affine and nonaffine contributions to the bulk modulus, and their dependence on the proximity to the unjamming point, can qualitatively differ between different models that feature the same generic unjamming phenomenology. In canonical models of unjamming we observe that the ratio of the nonaffine to total bulk moduli $B_{na}/B$ approaches a constant upon decompression, while in other, less well-studied models, it vanishes. We show that the vanishing of $B_{na}/B$ in non-canonical models stems from the emergence of an invariance of net (zero) forces on the constituent particles to compressive strains at the onset of unjamming. We provide a theoretical scaling analysis that fully explains our numerical observations, and allows to predict the scaling behavior of $B_{na}/B$ upon unjamming, given the functional form of the pairwise interaction potential.

\end{abstract}

\maketitle

\section{introduction}
Many disordered substances such as foams, emulsions, suspensions, and granular materials can jam into a solid-like state, or rather display fluid-like behavior, depending on their confining volume or pressure \cite{epitomeofdisorder2003,overviewpaperHecke,liu_review}. This diverse class of systems, that exhibit a continuous but abrupt transition between solid and fluid states by decompression, are often modeled theoretically and computationally by assemblies of soft repulsive particles \cite{epitomeofdisorder2003}. Using these models, it has been shown that the unjamming transition is accompanied by a number of intriguing phenomena, such as the emergence of an excess of low frequency vibrational modes \citep{epitomeofdisorder2003}, diverging length scales \citep{ellenbroek2006,lerner2014breakdown,Lerner2018} and scaling laws of elastic moduli \citep{epitomeofdisorder2003,ellenbroek2009non}. One of the hallmarks of the unjamming transition of soft sphere packings is the eventual loss of their shear rigidity --- reflected by the vanishing shear-to-bulk moduli ratio $G/B$ --- upon decompression \cite{ellenbroek2009non}. Mean-field approaches, such as effective medium theory \cite{EMT_01,eric_boson_peak_emt} or rigidity percolation of random networks \citep{thorpe_rigidity_percolation}, are unable to capture the self-organizational processes that determine the scaling behavior of $G/B$ \cite{ellenbroek2009non}. 

Previous work \cite{lutsko} has shown that \emph{athermal} elastic moduli of disordered solids consist of two contributions with different physical origins: an \emph{affine} term (also referred to often as the ``Born" term \cite{born1954dynamical}), which captures the stiffness of the material with respect to imposed affine deformations, and a \emph{nonaffine} term that accounts for additional (nonaffine) displacements of the internal degrees of freedom, that are required in order to preserve mechanical equilibrium under the imposed deformation. Thus, we decompose the shear and bulk moduli as
\begin{equation}\label{decomposition}
G = G_{a}-G_{na} \quad \quad B = B_{a} - B_{na},
\end{equation}
where the subscripts `$a$' and `$na$' stand for the affine and nonaffine terms, respectively. The aformentioned vanishing of $G/B$ upon unjamming is seen as the near cancellation of $G_{a}$ and $G_{na}$ as the unjamming point is approached, while the nonaffine term of the bulk modulus always remains significantly smaller than the affine term \cite{elasticmoduli,zaccone2014short}.

In this work we show that the relative smallness of the nonaffine term of the bulk modulus $B_{na}/B$ is non-universal across various model systems of purely repulsive soft spheres in two dimensions (2D), and focus on understanding its physical origin. The scaling properties of $B_{na}/B$ have been previously explained in terms of excluded volume correlations \citep{schlegel2016local,zaccone2014short}, and in terms of the behavior of states of self stress \citep{matthieu_thesis,ellenbroek2009non}. Here we put forward a different perspective on this problem; below we show that the relative smallness of $B_{na}$ stems from two key ingredients: $(i)$ the approximate proportionality between pairwise forces and pairwise stiffnesses, that can be directly inferred from the pairwise interaction potential, and $(ii)$ that the \emph{net} force acting on each particles vanishes --- the mechanical equilibrium condition, also invoked in the argumentation of  \citep{matthieu_thesis,ellenbroek2009non}.

In what follows we study the canonical unjamming model \cite{epitomeofdisorder2003}: soft spheres that interact via a $\varphi \sim \delta^n$ interaction, where $\delta$ denotes the overlap between neighboring spheres, and for most of our study we focus on $n\! =\! 2$ and $n\!=\! 3$. We find that, in these models, the ratio $B_{na}/B$ approaches a constant in the limit of zero pressure, at the onset of the unjamming transition. We also study two other models that feature the same loss of shear rigidity as observed in the canonical models, while at the same time exhibit \emph{qualitatively different} scaling behavior of the ratio $B_{na}/B$; upon approaching the unjamming point, $B_{na}/B$ gradually \emph{vanishes}, in contrast with its behavior in the canonical models. 

This paper is organized as follows; in Sect.~\ref{models_and_methods} we provide descriptions of the different models that we explore and the employed numerical methods. In Sect.~\ref{results} we present the results from our numerical simulations regarding the unjamming phenomenology of the different models, with particular focus devoted to the ratio $B_{na}/B$. In Sec.~\ref{discussion} we construct a string of scaling arguments, validated by our numerical simulations, that culminate in a scaling theory that fully explains the observed scaling laws of the ratio $B_{na}/B$ as unjamming is approached. In Sect.~\ref{summary} we summarize our findings and discuss future work.

\section{models, methods and observables}
\label{models_and_methods}

In this Section we describe the numerical models and methods employed in our work, and define the observables of interest. 

\subsection{Soft sphere models}
\begin{figure}[!ht]
\centering
\includegraphics[width = 0.4\textwidth]{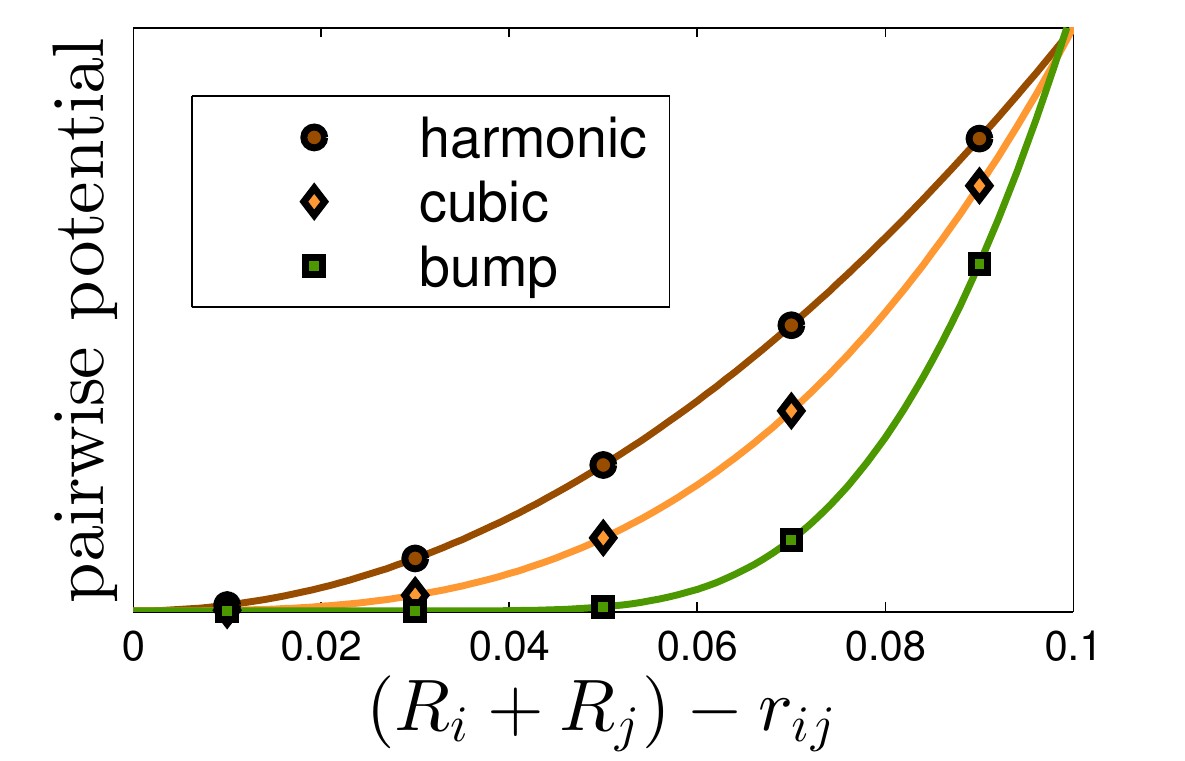}
\caption{\footnotesize Plotted are the harmonic, cubic and bump pairwise potentials, as a function of the overlap $(R_i\! + \! R_j)\! -\! r_{ij}$ of particles $i$ and $j$. Potentials have been factored for visualization purposes.}
\label{fig:potentials_fig}
\end{figure}

In this work we make use of four different models of soft, purely repulsive spheres in 2D. Disordered packings were generated for all models by a short high temperature equilibration, following by a minimization of the potential energy by means of the FIRE algorithm \cite{fire}. In all models except the EXP (introduced below) model, we incorporated the Berendsen barostat \cite{berendsen} into our minimization algorithm, allowing us to generate packings at any desired target pressures, as explained in \cite{swapjamming}. For the EXP model we employed 128-bit numerics for obtaining packings at extremely small densities. For all models, at all investigated state points, we employed 1000 independent packings of $N\!=\!1600$ particles. Plots of the employed pairwise potentials with a finite-range cutoff are presented in Fig.~\ref{fig:potentials_fig}. 


\subsubsection{Harmonic and cubic spheres}
We employed the canonical model of unjamming \cite{epitomeofdisorder2003}, in which spheres of radii $R_i$ and $R_j$ interact via the pairwise potential
\begin{equation}\label{canonical_potential}
\varphi_n(r_{ij},R_i,R_j) = \left\{\begin{array}{cc}\!\!\frac{\varepsilon}{n\lambda^n}\big((R_i\! +\! R_j) \!-\! r_{ij}\big)^n\!\!,&r_{ij}\!\le\! R_i\! + \! R_j\\0\quad\quad,\!\!&r_{ij}\!>\! R_i\! + \! R_j\end{array}\right.\!\!\!,
\end{equation}
with $\varepsilon$ and $\lambda$ denoting microscopic units of energy and length, respectively, and $r_{ij}$ is the pairwise distance between the centers of particles $i$ and $j$. In this work we mainly focus on models with $n\!=\!2$ and $n\!=\!3$, referred to in what follows as the \emph{harmonic} and \emph{cubic} models, respectively. We chose the radii of half the particles to be $7\lambda/5$, and the other half's radii is $\lambda$. This model undergoes an unjamming transition at a packing fraction $\phi\!\equiv\! V/V_p\!\approx\!0.84$ where $V\!=\! L^2$ denotes the volume (in 2D), and $V_p\!=\!\sum_i \pi R_i^2$ is the volume occupied by the particles. 


\subsubsection{Exponential spheres}
The third model we employed is a binary mixture of particles that interact via an exponentially-decaying pairwise potential
\begin{equation}
\varphi_{\text{\tiny EXP}}(r_{ij}) = \varepsilon_{ij}e^{-r_{ij}/\ell_{ij}}\,,
\end{equation}
where the parameters $\ell_{ij}$ and $\epsilon_{ij}$ depend on the species of the pair $i,j$, and can be found in \cite{eigen_paper}. As shown in \cite{eigen_paper}, this model undergoes an unjamming transition at vanishing densities $\rho\!\equiv\! N/V\!\to\!0$. We refer to this model in what follows as the EXP model. 

\subsubsection{The ``bump" model}
The fourth model we employed is a binary mixture in which pairs of particles interact via the ``bump" pairwise potential
\begin{equation}
\varphi_{\text{\tiny BUMP}}(r_{ij},R_i,R_j) =  \left\{\begin{array}{cc}\!\!\varepsilon e^{-\left(1-\left(\frac{r_{ij}}{R_i+R_j}\right)^{2}\right)^{-1}}\!\!\!\!\!,&r_{ij}\!<\! R_i\! + \! R_j\\0\quad\quad,\!\!&r_{ij}\!>\! R_i\! + \! R_j\end{array}\right.\!\!\!,
\end{equation}
where $\varepsilon$ denotes the microscopic units of energy. As in the case for the harmonic and cubic systems, half of the particles are of size $7\lambda/5$, and the other half of size $\lambda$, where $\lambda$ denotes the microscopic units of length. We refer to this model as the ``bump" model. As far as we know, it has not been studied before in the context of the unjamming transition. 

\subsection{Observables}

Athermal elastic moduli were calculated following the formalism put forward in \cite{lutsko}; the key principle in deriving expressions for athermal elastic modulus is that the system remains in mechanical equilibrium under imposed deformations. We employed the definitions 
\begin{equation}\label{foo06}
G \equiv \frac{1}{V}\frac{d^2U}{d\gamma^2}\quad\mbox{and}\quad B \equiv \frac{1}{V}\frac{d^2U}{d \eta^2}
\end{equation}
for the shear and bulk modulus, respectively, where $U$ is the potential energy, $V$ is the volume, $\gamma$ is the simple shear strain and $\eta$ is the expansive strain. The latter two parametrize the imposed affine transformation of coordinates $\xv\!\to\! H\!\cdot\!\xv$ in 2D as
\begin{equation}
H = \left( \begin{array}{cc}1+\eta&\gamma\\0&1+\eta\end{array}\right)\,.
\end{equation}
Using this transformation, the strain tensor $\epsilon$ is given by
\begin{equation}
\epsilon = \frac{1}{2}\left( H^T\cdot H - {\cal I}\right) = \frac{1}{2}\left( \begin{array}{cc}2\eta + \eta^2&\gamma + \gamma\eta\\\gamma + \gamma\eta&2\eta + \eta^2 +\gamma^2\end{array}\right)\,,
\end{equation}
where ${\cal I}$ represents the identity tensor. In the athermal limit the potential energy density variation $\delta U/V\!\equiv\!\big(U(\epsilon)\!-\!U(0)\big)/V$ can be approximated by a Taylor expansion in terms of the strain tensor $\epsilon$ and the general elastic coefficients $C_{\kappa\chi}$ and $C_{\kappa\chi\theta\tau}$ of the form
\begin{equation}
\delta U/V \simeq \sum_{\kappa \chi} C_{\kappa\chi}\epsilon_{\kappa\chi} + \sFrac{1}{2}\sum_{\kappa\chi\theta\tau}C_{\kappa\chi\theta\tau}\epsilon_{\kappa\chi}\epsilon_{\theta\tau}\,.
\end{equation}
In terms of the general coefficients $C_{\kappa\chi}$ and $C_{\kappa\chi\theta\tau}$, our definitions of shear and bulk moduli given by Eq.~(\ref{foo06}) read
\begin{equation}
G = C_{yy} + C_{xyxy}\,,
\end{equation}
and 
\begin{equation}
B = C_{xxxx} + C_{yyyy} + 2C_{xxyy}\,.
\end{equation}
We note importantly that our definition of the bulk modulus $B$ as given by Eq.~(\ref{foo06}), chosen for the sake of simplicity, slightly differs from the conventional definition of the bulk modulus $K\!\equiv\! -V\,dp/dV$, with $p\!\equiv\!-dU/dV$ denoting the pressure. The two definitions are related via $B\!=\!(K\!-\! p)\dbar^2$, where $\dbar$ denotes the spatial dimension. As we shall show below, in the unjamming limit $p/B\!\to\!0$, then the two definitions agree (up to an unimportant factor of $\dbar^2$).

In our models the potential energy is given by a sum over pairwise radially-symmetric interactions, namely $U\!=\!\sum_{i<j}\varphi_{ij}(r_{ij})$; the atomistic expression for the bulk modulus writes
\begin{equation}
B = \frac{1}{V}\bigg(\sum_{i<j}\varphi''_{ij}r^{2}_{ij} - \sum_{k,l}\Xiv_{k}\cdot\calBold{M}^{-1}_{kl}\cdot\Xiv_{l}\bigg) 
\end{equation}
where $\varphi''_{ij}\!\equiv\!\partial^2\varphi_{ij}/\partial r_{ij}^2$, $\calBold{M}_{ij}\!\equiv\!\frac{\partial^2U}{\partial\xv_i\partial\xv_j}$, and 
\begin{equation}\label{sum_over_dipoles}
\Xiv_{k} \equiv \frac{\partial^2U}{\partial\eta\partial\xv} = \sum_{i<j} \varphi''_{ij}r_{ij} \dv^{\,ij}_{k}
\end{equation} 
is the force (linear) response to an imposed expansive strain, written as a weighted sum over dipole vectors
\begin{equation}\label{define_dipole}
\dv^{\,ij}_{k} \equiv \frac{\partial r_{ij}}{\partial\xv_k} = (\delta_{jk}-\delta_{ik})\nv_{ij}\,,
\end{equation}
with $\nv_{ij}$ the unit vector pointing from particle $i$ towards particle $j$. Following the decomposition of $B$ as spelled out in Eq.~(\ref{decomposition}), we identify the affine and nonaffine contributions to the bulk modulus as
\begin{equation}\label{eq:bulkmodulus}
B_a\equiv \frac{1}{V}\sum_{i<j}\varphi''_{ij}r^{2}_{ij}\,, \ \ \mbox{and} \ \ B_{na}\equiv\frac{1}{V}\sum_{k,l}\Xiv_{k}\cdot\calBold{M}^{-1}_{kl}\cdot\Xiv_{l}\,.
\end{equation}

\begin{figure*}[!ht]
\centering
\includegraphics[width = \textwidth]{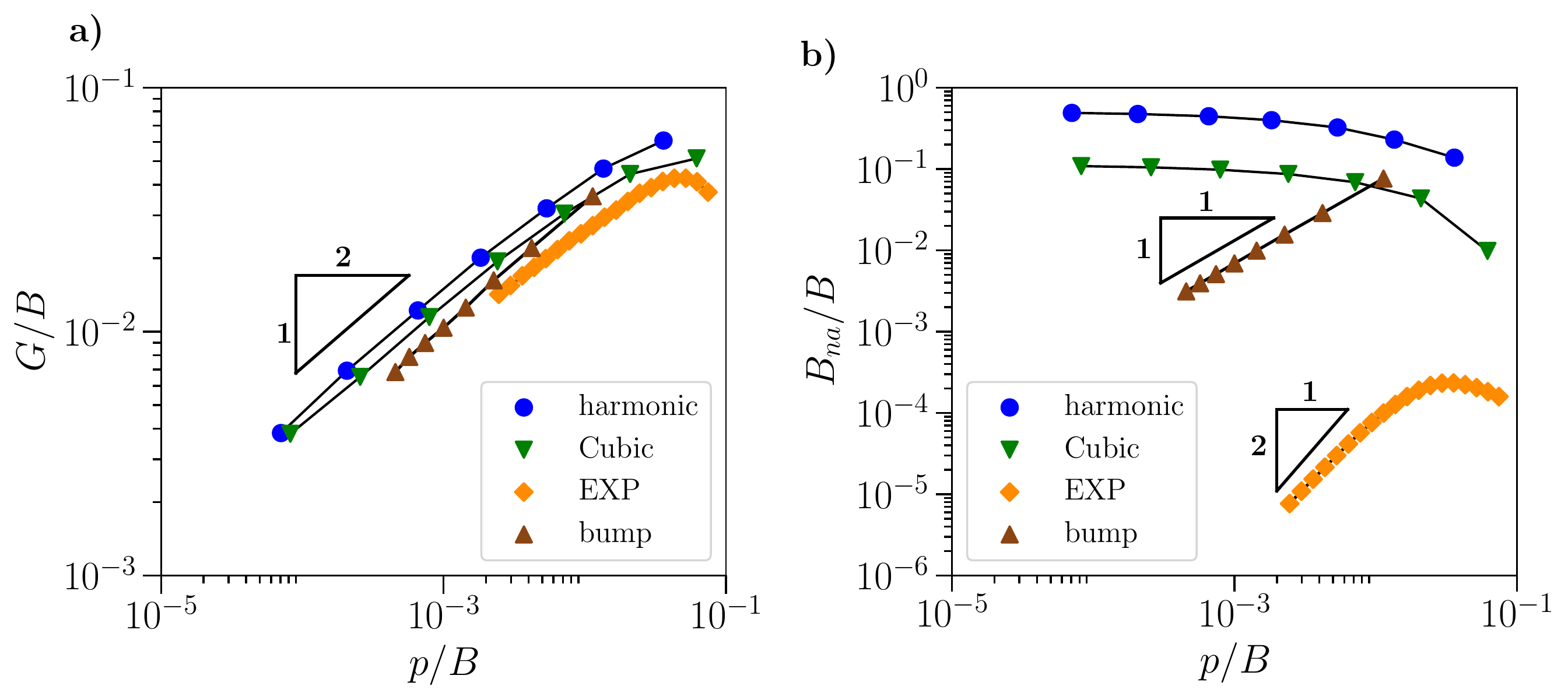}
\caption{\footnotesize (a) The shear-to-bulk modulus ratio, $G/B$, versus $p/B$ for all investigated models of soft repulsive spheres. The scaling $G/B\!\sim\!\sqrt{p/B}$ appears to universaly hold across all studied models. (b) The relative nonaffine contribution to the bulk modulus, $B_{na}/B$, plotted against the key dimensionless control parameter $p/B$. The harmonic and cubic system show a saturation as $p/B\!\to\!0$, while, in stark contrast with these canonical models, the EXP and bump models feature a vanishing $B_{na}/B$ as $p/B\!\to\!0$, with \emph{different} scaling laws.}
\label{fig:mu_over_B}
\end{figure*}

\section{Nonuniversality of nonaffinity under compressive strain}
\label{results}

One of the key characteristics of the unjamming transition in soft sphere packings is the loss of shear rigidity captured by the vanishing of the shear-to-bulk modulus ratio $G/B$ upon approaching the transition. In Fig.~\ref{fig:mu_over_B}a we plot the ratio $G/B$ as a function of $p/B$, the latter will serve in what follows as our central dimensionless control parameter of the unjamming transition, that occurs when $p/B\!\to\!0$. Our data for all four investigated models are consistent with previous results for the canonical models \cite{overviewpaperHecke,liu_review}, namely that $G/B\!\sim\!\sqrt{p/B}$. In previous work we have also observed this scaling for soft sphere packings with inverse-power-law pairwise interactions \cite{eigen_paper}, strongly supporting its universality.

In Fig.~\ref{fig:mu_over_B}b we plot the ratio of the nonaffine to total bulk modulus $B_{na}/B$, against $p/B$; while the scaling of the shear rigidity $G/B\!\sim\!\sqrt{p/B}$ appears to be universal for soft sphere packings, $B_{na}/B$ displays very different behavior between our different models: the canonical models (harmonic and cubic) show an initial slight \emph{increase} of $B_{na}/B$ upon decompression, after which the ratio saturates to an $n$-dependent constant upon approached the unjamming point as $p/B\!\to\!0$. 

Remarkably, the EXP and bump models show opposite trends; as $p/B\!\to\!0$ the fraction $B_{na}/B$ \emph{vanishes} instead of converging to a constant, with scalings consistent with $\sim\!(p/B)^{2}$ for the EXP model, and $\sim\! p/B$ for the bump model. Moreover, at the same high $p/B$ the nonaffine term is relatively much smaller for the EXP model compared to the canonical models, by several orders of magnitude. 

In the next Section we will build a string of scaling arguments and present supporting data from our numerical tests, that will fully explain the various scalings of $B_{na}/B$ vs.~$p/B$ as shown in Fig.~\ref{fig:mu_over_B}.

\section{Scaling theory for $B_{na}/B$}
\label{discussion}

\subsection{What makes $B_{na}/B$ small?}
\label{small}

To explain the scaling of $B_{na}/B$ upon approaching the unjamming transition, we first go back to Eq.~(\ref{eq:bulkmodulus}) for the nonaffine term of the bulk modulus, $B_{na}$; we define $\Xi\!\equiv\!|\Xiv|$ and $\hat{\Xiv}\!\equiv\!\Xiv/\Xi$, then $B_{na}/B$ can be trivially written in the form 
\begin{equation}\label{foo02}
\frac{B_{na}}{B} =\frac{\Xi^{2}}{VB}\;\hat{\Xiv}\cdot \calBold{M}^{-1}\cdot \hat{\Xiv}\,.
\end{equation}
This form makes clear that the smallness of $B_{na}/B$ can stem either from the smallness of $\Xi$ alone, or by weak coupling between the eigenfunctions $\mathBold{\Psi}_\omega$ of $\calBold{M}$ and $\hat{\Xiv}$.

Let us therefore first consider the quadratic form $\hat{\Xiv}\!\cdot\!\calBold{M}^{-1}\!\cdot\!\hat{\Xiv}$; using the spectral decomposition of the dynamical matrix $\calBold{M}\!=\!\sum_\omega \omega^2\mathBold{\Psi}_\omega\!\otimes\!\mathBold{\Psi}_\omega$, it can be written~as 
\begin{equation}\label{foo00}
\hat{\Xiv}\cdot\calBold{M}^{-1}\cdot\hat{\Xiv} = \sum_{\omega} \frac{\big(\mathBold{\Psi}_\omega\cdot\hat{\Xiv}\big)^2}{\omega^2}\,.
\end{equation}
It has been shown \cite{matthieu_thesis,elasticmoduli} that modes near unjamming project onto pairwise directors as $\dv\cdot\Psi_\omega\!\sim\!\omega/\omega_0$, where $\dv$ is a local dipole vector as defined in Eq.~(\ref{define_dipole}), and $\omega_0\!\sim\!\sqrt{\bar{\varphi}''}$ is the square root of a characteristic pairwise stiffness. Since $\Xiv$ is a weighted sum of dipoles (see Eq.~(\ref{sum_over_dipoles})), and is normalized (i.e.~$\hat{\Xiv}\!\cdot\!\hat{\Xiv}\!=\!1$), we expect that $\big(\mathBold{\Psi}_\omega\cdot\hat{\Xiv}\big)^2\!\sim\!(\omega/\omega_0)^2/V$. The sum in Eq.~(\ref{foo00}) thus follows
\begin{equation}\label{foo01}
\sum_{\omega} \frac{\big(\mathBold{\Psi}_\omega\cdot\hat{\Xiv}\big)^2}{\omega^2} \sim\frac{1}{\omega_0^2}\int\limits_{\omega_*}^{\omega_0}D(\omega)d\omega\,,
\end{equation}
where $D(\omega)$ is the density of states, and $\omega_*$ is the well-studied frequency scale above which a plateau appears in $D(\omega)$ in unjamming systems \cite{epitomeofdisorder2003,matthieu_thesis,elasticmoduli,overviewpaperHecke,liu_review}, known to universally follow $\omega_*/\omega_0\!\sim\!\sqrt{p/B}$ in decompressed soft sphere packings near unjamming. 

\begin{figure}[!ht]
\centering
\includegraphics[width = 0.5\textwidth]{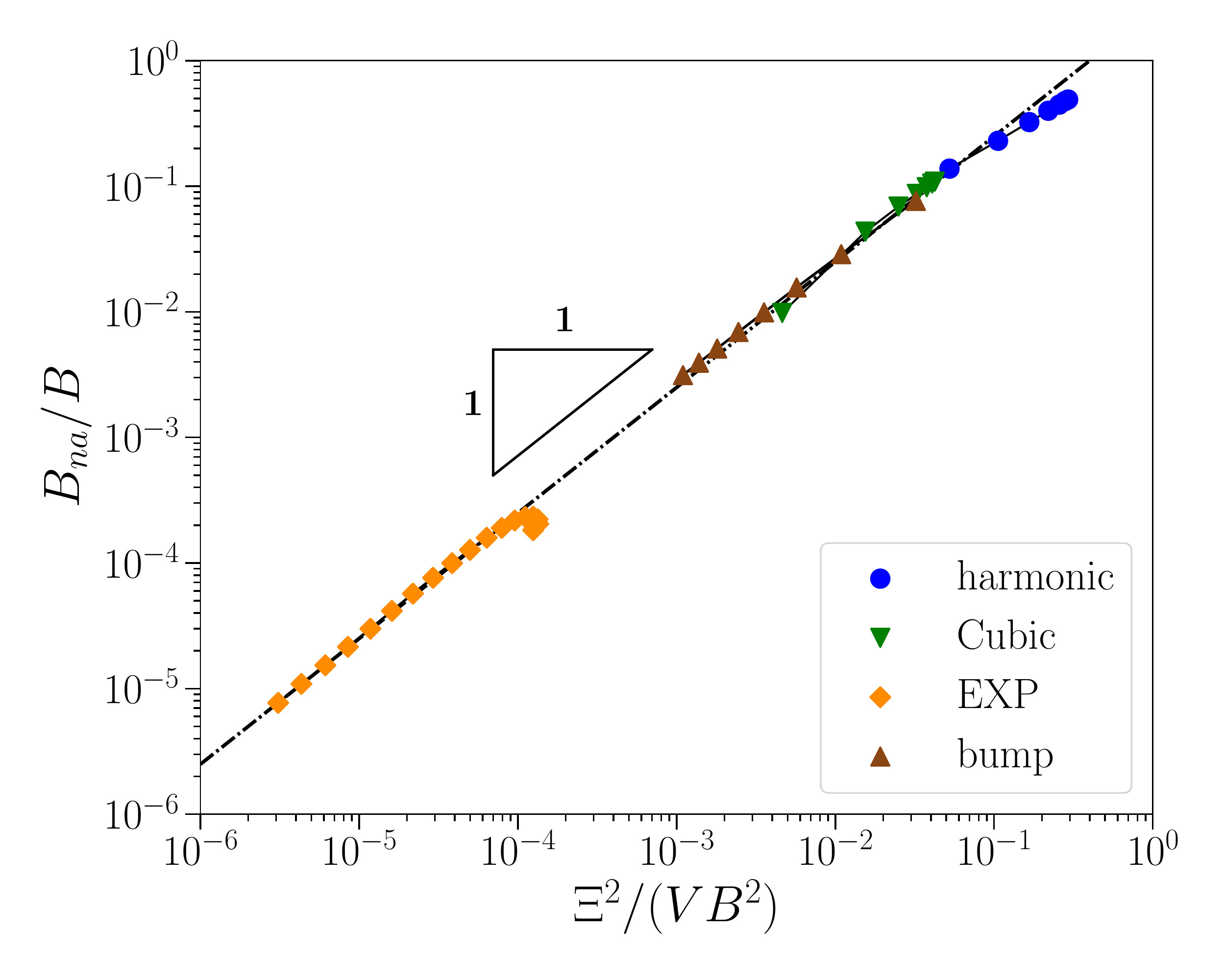}
\caption{\footnotesize $B_{na}/B$ versus $\Xi^{2}/(VB^{2})$ for all investigated models. The dashed line represents the fit $B_{na}/B =5\;\Xi^{2}/(2\;VB^{2})$, validating Eq.~(\ref{foo04}). }
\label{fig:scaling_Bna_over_B_Xi2_over_VB2}
\end{figure}

The main support of the density of states $D(\omega)$ near unjamming is between $\omega_*$ and $\omega_0\!\gg\!\omega_*$, implying that, to leading order $\int_{\omega_*}^{\omega_0}D(\omega)d\omega\sim {\cal O}(1)$. Finally, as long as $B\!\gg\! B_{na}$ then in 2D $\omega_0\!\sim\!\sqrt{B}$ (ignoring the unimportant units of mass) and therefore Eqs.~(\ref{foo00}) and (\ref{foo01}) imply that
\begin{equation}\label{foo03}
\hat{\Xiv}\cdot \calBold{M}^{-1}\cdot \hat{\Xiv} \sim 1/B\,.
\end{equation}
Combining Eqs.~(\ref{foo03}) and (\ref{foo02}), we conclude that
\begin{equation}\label{foo04}
\frac{B_{na}}{B} \sim \frac{\Xi^{2}}{VB^2}\,.
\end{equation}
In Fig.~\ref{fig:scaling_Bna_over_B_Xi2_over_VB2} we test this prediction; we find that not only are $B_{na}/B$ and $\Xi^2/(VB^2)$ proportional to each other, but they also share the \emph{same} proportionality constant across all investigated systems, which we find to be $5/2$. The conclusion from this analysis is that the smallness of $B_{na}/B$ must stem from the smallness of the rescaled compression-induced forces $\Xi/(B\sqrt{V})$.

\subsection{The roles of mechanical equilibrium and pairwise potential}

We continue the discussion by explaining how the condition of mechanical equilibrium, together with the functional form of the pairwise potential, affect the magnitude $\Xi$ of compression-induced forces $\Xiv$, and therefore control the scaling of $B_{na}/B$ near unjamming. Comparing the mechanical equilibrium equation and the definition of $\Xiv$ (see Eq.~\ref{eq:bulkmodulus}):
\begin{equation}
\mathBold{F}_{k} \equiv -\frac{\partial U}{\partial\xv_k} = - \sum_{i<j} \varphi_{ij}'\dv^{\,ij}_{k} = \zerovector\,, \quad \Xiv_{k} = \sum_{i<j} \varphi_{ij}'' r_{ij}\dv^{\,ij}_{k},
\end{equation}
we see that they have similar structures; both forces $\mathBold{F}$ and $\Xiv$ are given by a weighted sum of dipoles $\dv$, where the weight factors are the pairwise forces $\varphi'$ in the case of $\mathBold{F}$, and are the products of the pairwise stiffnesses and distances $\varphi'' r$ for the case of $\Xiv$. Therefore, in the special case $\varphi'\! =\! c\,\varphi''r$ --- with a dimensionless proportionally coefficient $c$ that is \emph{independent} of the pairwise distance $r$, as holds for widely-employed inverse-power-law (IPL) pairwise potentials, of the form $\varphi_{\mbox{\tiny IPL}}\!\sim\! r^{-\beta}$ \cite{eigen_paper} --- we expect $\Xiv\!=\!\zerovector$ by virtue of mechanical equilibrium.

For other pairwise potentials different from $\varphi_{\mbox{\tiny IPL}}$,
\begin{equation}
c(r)\equiv\varphi'/\varphi''r
\end{equation} 
is generally a \emph{function} of the pairwise distance $r$, and will thus fluctuate between different pairs of interacting particles (due to fluctuations of pairwise distances), resulting in a nonzero $\Xiv$. Nevertheless, when approaching the unjamming point, spatial fluctuations in pairwise distances between interacting particles are known to decrease \cite{epitomeofdisorder2003,jacquin2011microscopic}, in some cases (depending on the form of the interacting potential, see below) giving rise to an \emph{approximate} proportionality between $\varphi'$ and $\varphi''r$, that render $\Xi$ small.

\begin{figure}[!ht]
\centering
\includegraphics[width = 0.48\textwidth]{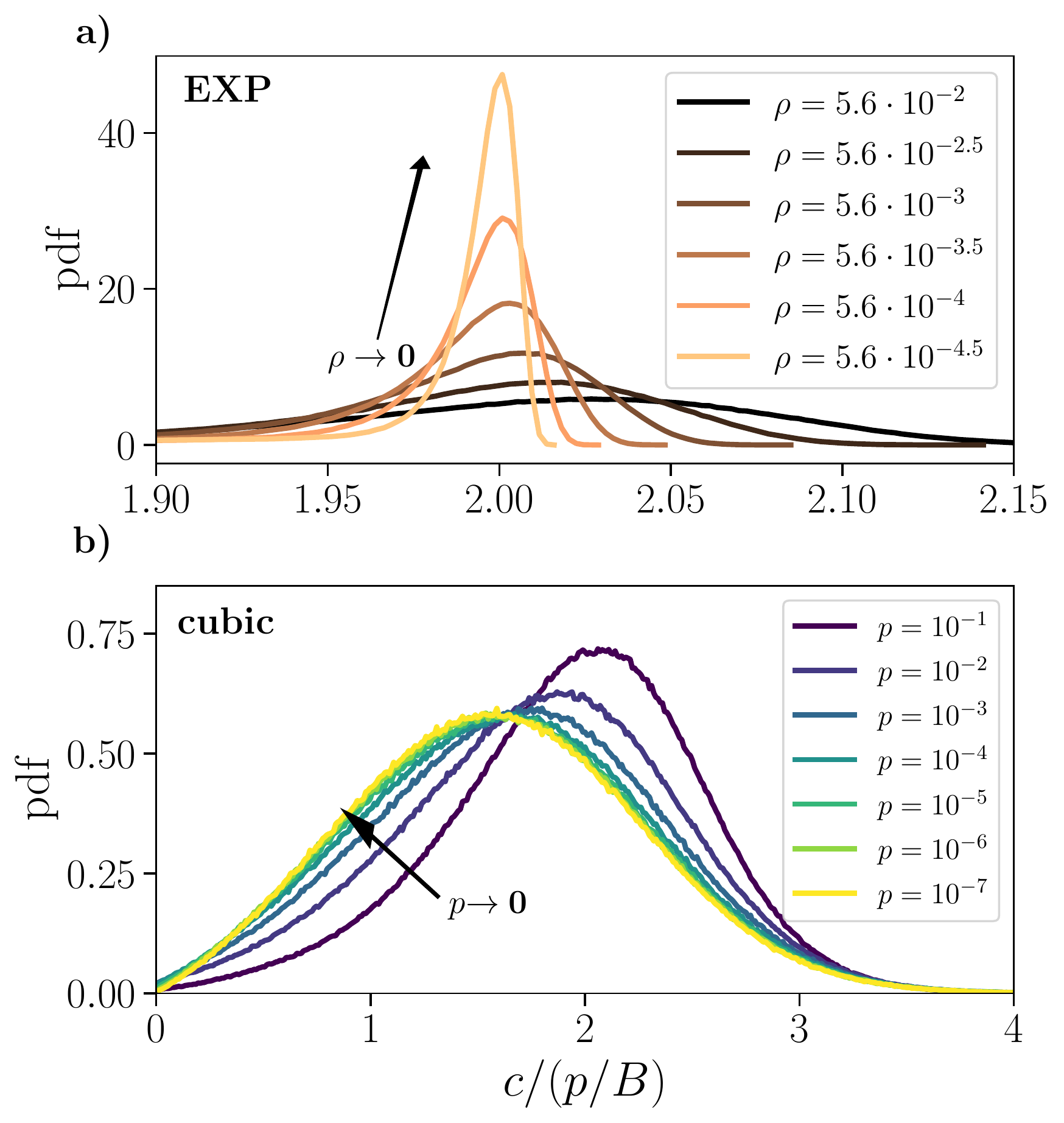}
\caption{\footnotesize Probability distribution functions of the proportionality coefficients $c\!\equiv\!\varphi'/\varphi''r$ for the EXP (panel (a)) and cubic (panel (b)) models, plotted against the rescaled variables $c/(p/B)$. We see a clear narrowing of the distributions in the EXP model, whereas they approach a limit form in the case of the cubic model. We note that for the EXP model, only interactions within the first coordination shell were considered in this analysis.}
\label{fig:distribution_f_kr_cubic_EXP}
\end{figure}

The emergence of an approximate proportionality between $\varphi'$ and $\varphi''$ as unjamming is approached can be directly observed in our simulations. Before discussing data, we note that characteristic pairwise forces scale as $\bar{\varphi}'\!\sim\! p\,\bar{r}$ (in 2D) with $\bar{r}\!\equiv\!1/\sqrt{\rho}$ a characteristic pairwise distance, and, similarly,  characteristic pairwise stiffnesses scale as $\bar{\varphi}''\bar{r}\!\sim\! B\bar{r}$. We thus expect that characteristic proportionality coefficients $\bar{c}\!\equiv\!\bar{\varphi}'/\bar{\varphi}''\bar{r}\!\sim\! p/B$. This expectation is validated in Fig.~\ref{fig:distribution_f_kr_cubic_EXP}, in which the probability distributions $p(c)$ are plotted against the rescaled variable $c/(p/B)$, for the EXP (panel (a)) and cubic (panel (b)) models, as the unjamming transition is approached (i.e.~$\rho\!\to\!0$ and $p\!\to\!0$ for the two models, respectively). For the EXP model, there is a clear narrowing of the distribution, showing that correlations between $\varphi'$ and $\varphi''r$ increase as unjamming is approached, while for the cubic model the distributions approach a limit form as $p\!\to\! 0$. The bump and harmonic models show similar behavior (i.e.~the distributions narrow for the bump model, but approach a limit form for the harmonic model) and are therefore not shown here. 

\begin{figure*}[!ht]
\centering
\includegraphics[width = 0.85\textwidth]{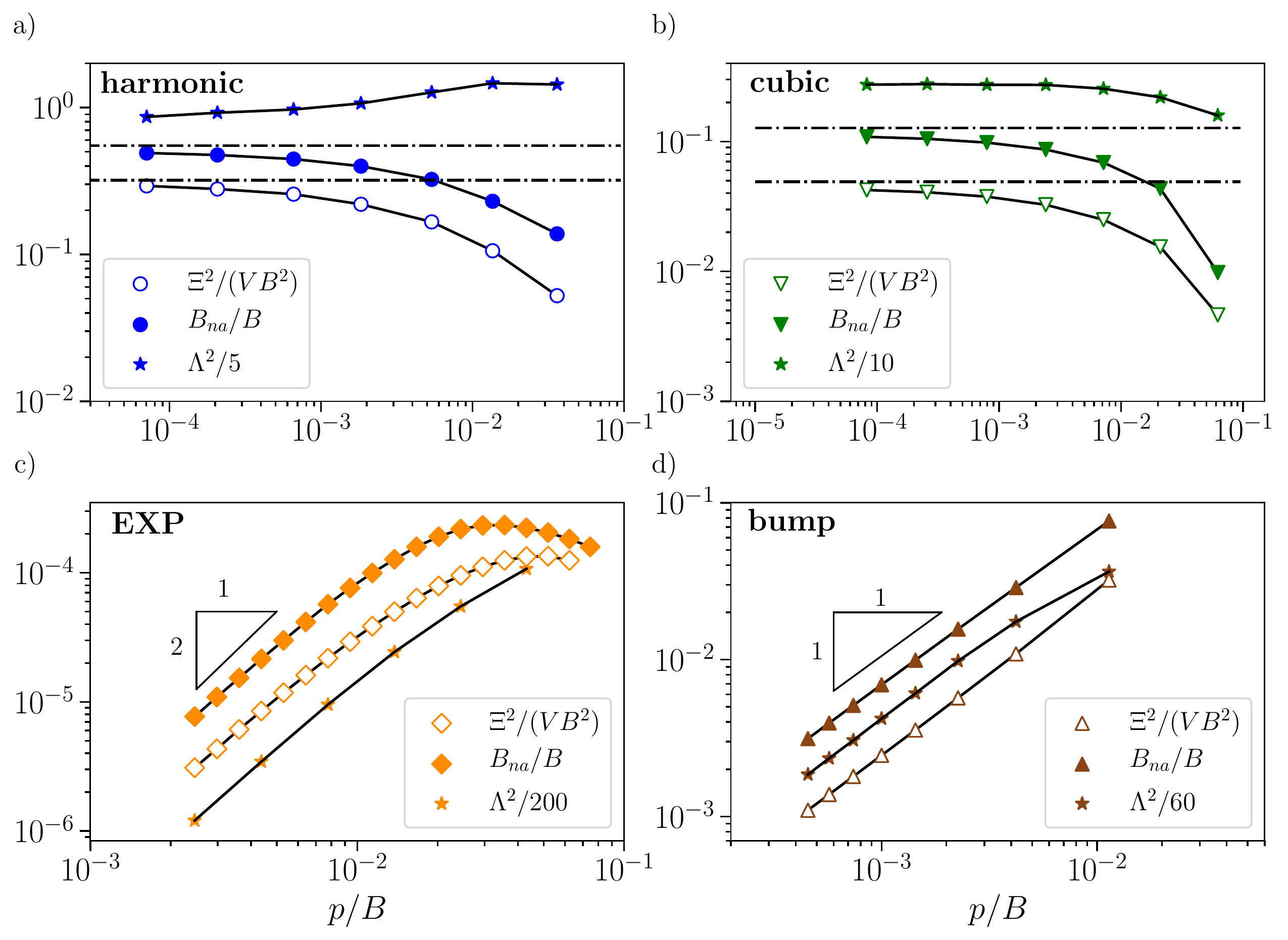}
\caption{\footnotesize Plots of $B_{na}/B$ versus $p/B$ (closed symbols) and $\Xi^{2}/(VB^{2})$ vs $p/B$ (open symbols), plus the squared full width at half maximum ($\Lambda^{2}$) of the distributions of $c/(p/B)$ (stars), see Fig.~\ref{fig:distribution_f_kr_cubic_EXP} for examples of these distributions. Panel (a) and (b) show the harmonic and cubic interactions, with a putative line of constant $B_{na}/B$. Panel (c) and (d) show the EXP and bump models, respectively. The data make clear that all three quantities; $B_{na}/B$, $\Xi^{2}/(VB^{2})$ and $\Lambda^{2}$ all share the same scaling (per model) with respect to $p/B$.}
\label{fig:Bna_over_B}
\end{figure*}

To quantify the reduction of relative fluctuations of the proportionality coefficients $c$ near unjamming, we determined the full width at half maximum (denoted in what follows by $\Lambda$) of their distributions, for all investigated models and pressures. The modes of the distributions are close to the value two, which simply originates from the factor $1/\dbar$ in the expression for the pressure. In Fig.~\ref{fig:Bna_over_B} we show that $\Lambda^{2}$ has the same scaling as $B_{na}/B$ in terms of $p/B$, supporting that it is controlled by the reduction of fluctuations of the proportionality coefficients $c$.

\subsection{Scaling theory for $B_{na}/B$}

To deduce the scaling of $B_{na}/B$ in terms of $p/B$ directly from the functional form of the pairwise potential, let us first express the compression-induced forces $\Xiv$ --- that we have shown to control $B_{na}/B$ in Sect.~\ref{small} --- in terms of the proportionality coefficients $c\!\equiv\!\varphi'/\varphi''r$ as
\begin{eqnarray}
\Xiv_{k} & = &\sum_{i<j} \varphi_{ij}'' r_{ij}\dv^{\,ij}_{k} = \sum_{i<j} c^{-1}_{ij} \varphi'_{ij}\dv^{\,ij}_{k} \nonumber \\ 
& =&  \sum_{i<j} \left(c^{-1}_{ij} - \langle c^{-1}\rangle + \langle c^{-1}\rangle\right)\varphi'_{ij}\dv^{\,ij}_{k} \nonumber \\ 
& \simeq & \frac{dc^{-1}}{dr}\sum_{i<j} \big(r_{ij} - \langle r \rangle \big)\, \varphi'_{ij}\dv^{\,ij}_{k}\,, \label{foo05}
\end{eqnarray}
assuming fluctuations of the proportionality coefficients $c$ around their mean are small, as seen in Fig.~\ref{fig:distribution_f_kr_cubic_EXP} and Fig.~\ref{fig:Bna_over_B}. The procedure followed in Eq.~(\ref{foo05}) is akin to projecting out the so-called `states of self stress' \cite{Tighe_2011} from the weighted sum over dipoles, which bears similarity with the arguments of \cite{matthieu_thesis} regarding the behavior of the bulk modulus near unjamming. Using that $\frac{dc^{-1}}{dr}\!\propto\!\frac{1}{c^2}\frac{dc}{dr}$ and denoting $\Delta r_{ij}\!\equiv\! r_{ij}\!-\!\langle r \rangle$, the squared compression-induced force follows
\begin{equation}\label{foo07}
\Xi^2 = \sum_k \Xiv_k\cdot\Xiv_k \sim \frac{1}{c^4}\bigg(\frac{dc}{dr}\bigg)^2 \sum_{i<j}\left( \Delta r_{ij} \varphi_{ij}'  \right)^2\,,
\end{equation}
where we neglected subleading off-diagonal contributions. 

\begin{figure}[!ht]
\centering
\includegraphics[width = 0.5\textwidth]{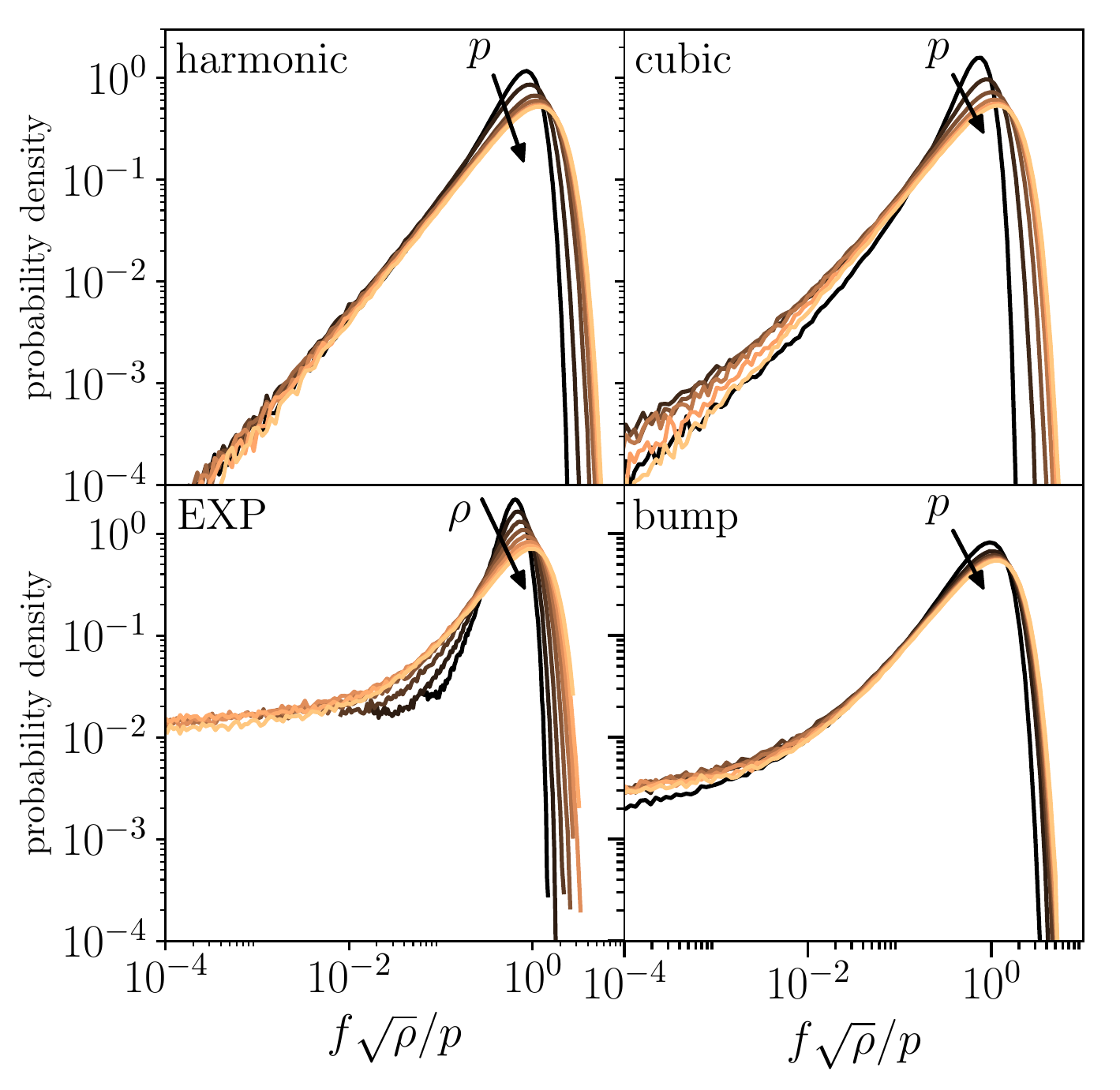}
\caption{\footnotesize Rescaled force distributions for the different potentials. Clearly the distributions obtain a constant shape as the unjamming transition is approached. This can be used to determine the scaling of $\Delta r$ as a function of the control parameter.  }
\label{fig:force_distributions}
\end{figure}

At this point we exploit the observation that pairwise force fluctuations scale with the mean pairwise force, namely $\varphi'\!\sim\!\Delta\varphi'\!\equiv\!\varphi' \!-\! \langle\varphi'\rangle$, as demonstrated for all four investigated models in Fig.~\ref{fig:force_distributions}. Since $\Delta \varphi' \!\sim\! \varphi'' \Delta r$, then as long as $\varphi''\!\to\!0$ near unjamming, $\Delta \varphi'\!\ll\!\Delta r$ (and $\Delta \varphi'\!\sim\!\Delta r$ in the harmonic model for which $\varphi''$ is constant). We can thus neglect fluctuations of $\varphi'_{ij}$ in Eq.~(\ref{foo07}), and only keep the leading order contribution, that reads
\begin{equation}
\Xi^2 \sim \frac{1}{c^4}\bigg(\frac{dc}{dr}\bigg)^2 (\bar{\varphi}')^2\sum_{i<j}\big(\Delta r_{ij}\big)^2\,.
\end{equation}
Since $\bar{c}\!\sim\! p/B$, $\bar{\varphi}'\!\sim\! p\,\bar{r}$, and $N/V\!\sim\!1/\bar{r}^2$ (in 2D), together with Eq.~(\ref{foo04}) we obtain
\begin{equation}
\frac{B_{na}}{B} \sim \frac{\Xi^2}{VB^2} \sim \frac{B^2}{p^2}\bigg(\frac{dc}{dr}\bigg)^2 \langle (\Delta r)^2 \rangle\,.
\end{equation}
As discussed above, characteristic fluctuations of pairwise distances follows $\Delta r\!\sim\!\Delta\varphi'/\varphi''\!\sim\!\varphi'/\varphi''\!\sim\! \bar{r}\,p/B$, which leads us to our key result for 2D packings
\begin{equation}\label{eq:scaling_of_nonaffinebulk}
\frac{B_{na}}{B} \sim \bigg(r\frac{dc}{dr}\bigg)^2 = \left(1-\frac{\varphi'(\varphi'''r + \varphi'')}{(\varphi'')^2r}\right)^2\,.
\end{equation}
Our observation that $\Lambda^2\!\sim\! B_{na}/B$, as seen in Fig.~\ref{fig:Bna_over_B}, is now readily explained; $\Lambda$ should scale as a characteristic fluctuation of $c$, rescaled by $p/B$, namely
\begin{equation}
\Lambda \sim \frac{\Delta c}{p/B}\equiv \frac{\sqrt{\langle(c - \langle c \rangle)^2\rangle}}{p/B} \sim \frac{\frac{dc}{dr}\Delta r}{p/B} \sim r\frac{dc}{dr}\sim \sqrt{\frac{B_{na}}{B}}\,,
\end{equation}
in agreement with our observation.

\subsection{Model-dependent scaling laws for $B_{na}/B$}
\begin{figure}[!h]
\centering
\includegraphics[width = 0.43\textwidth]{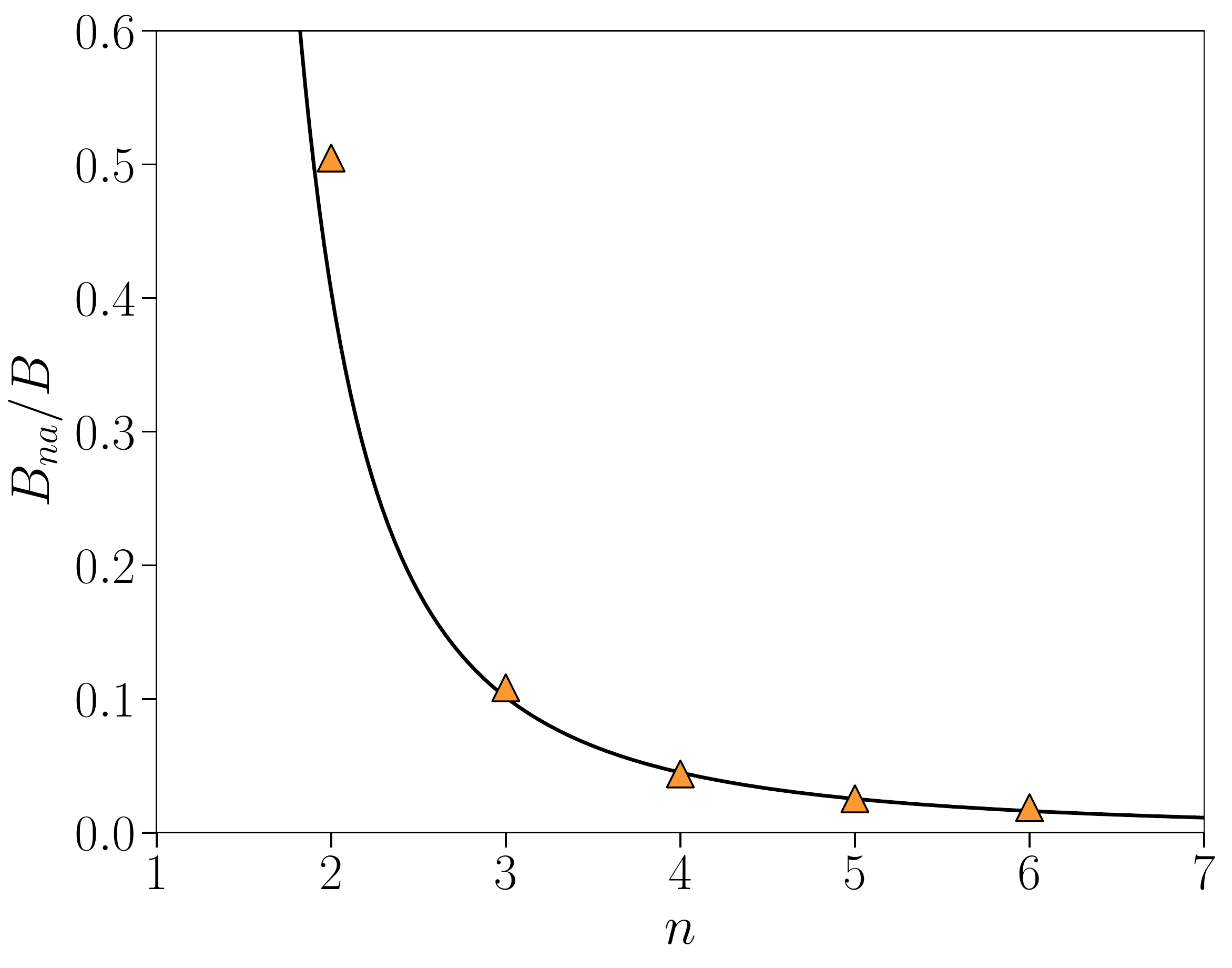}
\caption{\footnotesize $B_{na}/B$ at low pressures for the canonical models with different exponents $n$. The continuous line represents a fit to the form $1/(n-1)^{2}$ as predicted by Eq.~(\ref{foo08}).}
\label{fig:canonical_models_Bna_n}
\end{figure}

Now we can readily explain the measured scalings laws of $B_{na}/B$ in terms of $p/B$ for the different models, as presented in Fig.~\ref{fig:Bna_over_B}, using Eq.~(\ref{eq:scaling_of_nonaffinebulk}). The harmonic and cubic interactions employ a pairwise potential of the form given by Eq.~(\ref{canonical_potential}). Using Eq.~(\ref{eq:scaling_of_nonaffinebulk}) we see that near unjamming
\begin{equation}\label{foo08}
\frac{B_{na}}{B} \sim \frac{1}{(n-1)^{2}} \to \mbox{constant},
\end{equation}
in agreement with our data of Figs.~\ref{fig:mu_over_B} and~\ref{fig:Bna_over_B}. Interestingly, this also shows that the nonaffine contribution to the bulk modulus near unjamming decreases for higher exponents $n$. Fig.~\ref{fig:canonical_models_Bna_n} shows $B_{na}/B$ near unjamming, for systems with increasing $n$, confirming once more the predicted scaling. 

For the EXP model, we make use of the relation $p/B\!\sim\!\sqrt{\rho}$ as derived in \cite{eigen_paper} for the EXP model in 2D, to predict that
\begin{equation}
\frac{B_{na}}{B} \sim \frac{1}{\bar{r}^{2}} \sim \rho \sim \left(\frac{p}{B}\right)^{2},
\end{equation}
in agreement with our data of Figs.~\ref{fig:mu_over_B} and~\ref{fig:Bna_over_B}.

Finally, it can be shown that for the bump model near the unjamming transition $dc/dr\!\sim\!\sqrt{\varphi'/\varphi''r}$, leading to 
\begin{equation}
\frac{B_{na}}{B} \sim \frac{p}{B}\,,
\end{equation}
in agreement with our data of Figs.~\ref{fig:mu_over_B} and~\ref{fig:Bna_over_B}.

\section{Discussion and outlook}
\label{summary}

In this work we investigated the relative contribution of the nonaffine term to the bulk modulus, $B_{na}/B$, across four different models of soft sphere packings in 2D, near their respective unjamming points. We find that as the critical point is approached, the relative contribution of the nonaffine term to the bulk modulus is non-universal: it can either saturate to a constant or vanish, depending on the form of the pairwise interaction potential.

In order to explain the non-universality in the observed scaling laws, we first established that the smallness of $B_{na}/B$ stems from the relative smallness of the compression-induced forces $\Xiv\!\equiv\!\partial^2U/\partial\eta\partial\xv$, which are a key component of $B_{na}$ (see Eq.~(\ref{eq:bulkmodulus})). This was done by arguing that $B_{na}/B\!\sim\!\Xi^2/VB^2$, which was also verified numerically. Interestingly, we found that not only does this scaling relation hold for all investigated systems, but also that the dimensionless proportionality coefficient of this scaling law appears to be universal across all investigated models, suggesting that it might be amenable to an exact calculation in future work. 

In two of the four investigated models, namely the EXP and bump models, the compression-induced forces $\Xiv$ are shown to vanish as the unjamming point is approached. This implies that at the critical point the net (zero) force becomes invariant to compressive or expansive strains. We speculate that a close connection exists between this emergent property of the EXP and bump models, and the approximate `isomorph-invariance' put forward by Dyre and co-workers \cite{Dyre_2016,Andreas}. It has been argued that high-dimensional soft sphere packings are isomorph-invariant \cite{Maimbourg_2016}, which might have implications on the dimensionality-dependence of nonaffinity upon compression as studied here. The detailed investigation of these connections is left for future work. 

We next proceeded to construct a string of scaling arguments that lead to our key result given by Eq.~(\ref{eq:scaling_of_nonaffinebulk}), namely that $B_{na}/B\!\sim\!( r\,dc/dr )^2$, where $c(r)\!\equiv\!\varphi'/\varphi''r$ is the proportionality coefficient between the pairwise force $\varphi'$ and the product $\varphi''r$, where $\varphi''$ and $r$ are the pairwise stiffness and distance between interacting particles, respectively. We showed that the derived scaling relation fully explains the behavior of $B_{na}/B$ near unjamming in all four investigated models.

Our scaling argument is based on the observation, presented in Fig.~\ref{fig:force_distributions}, that the distribution of pairwise (contact) forces, rescaled by $p/\rho^{1/\dbar}$, universally assumes a finite width near the unjamming point, independent of model details, as also seen in hard sphere packings \cite{C3SM50515D}, and in colloidal glasses near jamming \cite{corwin_prl_2012}. At this point we cannot argue why this must always be the case. An interesting result of this observation, however, is that fluctuations in pairwise distances follow $\Delta r\!\sim\! \bar{r}p/B$, where $\bar{r}$ is a characteristic pairwise distance. In the EXP model $\bar{r}\!\sim\! B/p$ \cite{eigen_paper}, indicating that, in this particular model, pairwise fluctuations are independent of density. 

In two of the four investigated models, namely the EXP and bump models, the finite width of the rescaled pairwise forces translates to a vanishing width of the rescaled dimensionless proportionality coefficients $c/(p/B)\!\sim\! c/\langle c\rangle$. This implies, remarkably, that while pairwise force fluctuations scale as the mean pairwise force, namely $\Delta \varphi'\!\sim\!\langle\varphi'\rangle$, this is not the case for the proportionality coefficients $c$, whose fluctuations to mean ratio $\Delta c/\langle c\rangle\to0$ as unjamming is approached. In future work the implications of these reduced fluctuations might be explored. \vspace{0.1cm}

\acknowledgements
We would like to thank SURFsara for the support in using the Lisa Computer Cluster. This work is part of the Industrial Partnership Programme Hybrid Soft Materials that is carried out under an agreement between Unilever Research and Development B.V. and the Netherlands Organisation for Scientific Research (NWO). EL acknowledges support from the NWO (Vidi grant no.~680-47-554/3259).



\appendix

\end{document}